\documentclass[aps,prb,amsmath,amssymb,superscriptaddress,floatfix,showpacs,twocolumn]{revtex4}

\usepackage{graphicx}
\usepackage{epstopdf} 
\usepackage{hyperref}
\usepackage[usenames,dvipsnames]{color}
\usepackage{bm}

\begin{document}

\title{Multiboson spin-wave theory for Ba$_2$CoGe$_2$O$_7$, a spin-3/2 easy-plane N\'eel antiferromagnet with strong single-ion anisotropy}

\author{Judit Romh\'anyi}
\affiliation{Department of Physics, Budapest University of Technology and Economics, H--1111 Budapest, Budafoki \'ut 8, Hungary}

\author{Karlo Penc}
\affiliation{Institute for Solid State Physics and Optics, Wigner Research
Centre for Physics, Hungarian Academy of Sciences, H-1525 Budapest, P.O.B. 49, Hungary}
\affiliation{Department of Physics, Budapest University of Technology and Economics, H--1111 Budapest, Budafoki \'ut 8, Hungary}

\date{\today}

\begin{abstract} 
  We consider the square-lattice antiferromagnetic Heisenberg Hamiltonian extended with a single-ion axial anisotropy term as a minimal model for the  multiferroic Ba$_2$CoGe$_2$O$_7$. Developing a multiboson spin-wave theory, we investigate the dispersion of the spin excitations in this spin-3/2 system. As a consequence of a strong single-ion anisotropy, a stretching (longitudinal) spin-mode appears in the spectrum. The inelastic neutron scattering spectra of Zheludev {\it et al.} [Phys. Rev. B {\bf 68}, 024428 (2003)] are successfully reproduced by the low energy modes in the multiboson spin-wave theory, and we anticipate the appearance of the spin stretching modes at $\approx$4meV that can be identified using the calculated dynamical spin structure factors. We expect the appearance of spin stretching modes for any $S>1/2$ compound where the single-ion anisotropy is significant.
\end{abstract}
\maketitle

\section{Introduction and the model}

The spin excitations in isotropic Heisenberg models with long-range N\'eel order are well understood within the framework of the spin-wave theory, even in the extreme quantum spin-1/2 case on the square lattice, as exemplified by La$_2$CuO$_4$.\cite{INS_LaCu2O4_2001,INS_LaCu2O4_2010} In the spin-wave approach, the spin operators are transcribed using a Holstein-Primakoff boson, and expanding in $1/S$ around the classical $S\to\infty$ ground state, one gets a Hamiltonian quadratic in the bosonic operators that describes the spin excitations of the ordered state. 
 Interestingly, the spin-waves as described above do not take account of all the possibilities for larger spins. In $S>1/2$ quantum spin models higher order tensor interactions (like biquadratic exchange) and anisotropy terms (like axial anisotropy) may lead to multipolar or nematic ordering.\cite{Blume1969,ANDREEVandGRISHCHUK,nematicreview2011} When studying the dynamical properties of such systems, however, the conventional spin-wave approach fails, and one needs to introduce generalized bosonic operators related not only to the spin but also to higher order spin operators.\cite{Papanicolaou1984,Papanicolaou1985,ISI:A1985AXC5300038,Chubukov1990} Similarly, additional bosons are needed for spin systems with orbital degeneracy.\cite{PhysRevB.60.6584} Interacting spin multiplet systems have also been studied by an extended Holstein-Primakoff theory in which a boson is introduced for each energy level of the multiplet. Based on this approach the excitation spectra of Cu$_2$Fe$_2$Ge$_4$O$_{13}$ and Cu$_2$CdB$_2$O$_6$ have been theoretically reproduced and for both materials longitudinal modes have been reported that are related to the deformation of the spin wave-function on the magnetic ions, also leading to the reduction of the magnetic moment of the spin.\cite{Matsumoto2010}  A multiboson approach with spin-orbital coupled 27 single ion levels has been applied to the case of the La$_{1.5}$Sr$_{0.5}$CoO$_4$ \cite{Helme2009} and  La$_2$CoO$_4$ \cite{Babkevich2010}, where the Co ions are in octahedral environment.

Our work is inspired by the strongly anisotropic spin-3/2 multiferroic material, Ba$_2$CoGe$_2$O$_7$. This compound has tetragonal symmetry and can be characterized by layers of square lattices formed by the magnetic Co$^{2+}$ ions.\cite{Zheludev2003,Yi2008,Murakawa2010,Miyahara2011,Hutanu2011} As the neighboring cobalts are positioned in differently oriented tetrahedral environments, the unit cell contains two of these. Below the transition temperature $T_{\text{N\'eel}} = 6.7$ K, the magnetic moments are antiferromagnetically aligned in the plane of the  Co$^{2+}$ ions.\cite{Zheludev2003} Spin excitations have been studied by inelastic neutron scattering in Ref.~\onlinecite{Zheludev2003}, and the observed  dispersions were fitted using the conventional spin-wave theory based on large exchange anisotropy. Additional spin excitations at $\approx1$ THz energies were observed in light absorption spectra in Ref.~\onlinecite{Kezsmarki2011}. While these additional modes are beyond the conventional spin-wave theory, they were reproduced in an exact diagonalization study of small clusters by Miyahara and Furukawa\cite{Miyahara2011} using a Heisenberg Hamiltonian extended with a strong single-ion anisotropy of the $(S^z)^2$ form. Recently, we applied the aforementioned multiboson spin-wave theory to describe the modes observed in far-infrared absorption spectra in external magnetic field.\cite{FIR2012}
In the present paper, we aim to study the effect of the strong single-ion anisotropy on the zero-field spin-wave spectrum in the momentum space using the multiboson spin-wave approach, and to give predictions for the dispersion of the higher energy modes in the inelastic neutron scattering spectra.

Following Ref.~\onlinecite{Miyahara2011}, we consider the Hamiltonian 
\begin{eqnarray}
\mathcal{H}&=&J\sum_{\langle i,j\rangle}\left(\hat S^x_i \hat S^x_j+\hat S^y_i \hat S^y_j\right)+J_z \sum_{\langle i,j\rangle}\hat S^z_i \hat S^z_j+\nonumber\\
&{}&\Lambda\sum_i \left(\hat S^z_i\right)^2 \;,
\label{eq:Hamiltonian}
\end{eqnarray}
where $J$ and $J_z$ are the exchange couplings, $\Lambda$ is the strength of the single-ion anisotropy, the summation is over the $\langle i,j \rangle$ nearest neighbor sites, and the $z$-axis is perpendicular to the square lattice plane. In Ref.~\onlinecite{Miyahara2011}, a rather strong easy--plane anisotropy, 
$\Lambda/J\approx 8$ has been suggested in Ba$_2$CoGe$_2$O$_7$. Furthermore, a  Dzyaloshinskii-Moriya  interaction $\approx 0.04J$ has also been considered, which is in fact very small compared to the exchange coupling and anisotropy term, thus we omit it in the present study.

The paper is structured as follows: In Sec.~\ref{sec:vari} we shortly discuss the variational phase diagram of Hamiltonian~(\ref{eq:Hamiltonian}) as a function of easy-plane and exchange anisotropies. The variational wave function is then used in Sec.~\ref{sec:flavor} to construct a suitable boson basis and perform the  multiboson spin-wave approach. The spin-wave Hamiltonian is diagonalized numerically and analytically in the momentum space, and the behavior of the modes  for different exchange and anisotropy parameters is discussed. In Sec.~\ref{sec:Sq} we calculate the dynamical spin structure factor. These results are quantitatively compared to the inelastic neutron scattering measurements\cite{Zheludev2003} of the Ba$_2$CoGe$_2$O$_7$ in Sec.~\ref{sec:INS} and we draw conclusions in Sec.~\ref{sec:conclusions}. Finally, in Appendix the $\Lambda\to 0$ and $\Lambda\to \infty$ cases are discussed in detail.

\section{Variational ground state}\label{sec:vari}

\begin{figure}[tb]
\begin{center}
\includegraphics[width=8cm]{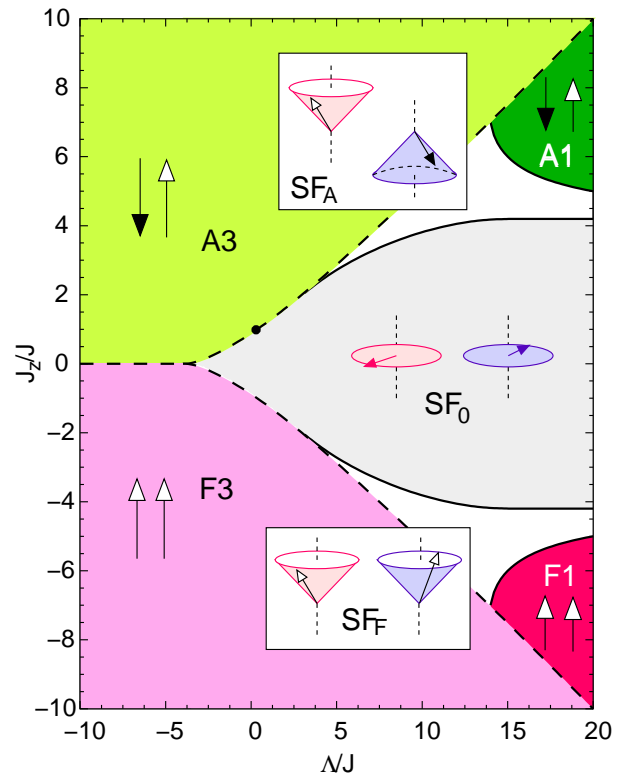}
\caption{(color online) Variational phase diagram for $h=0$ as the function of $\Lambda/J$ and $J_z/J$. Solid lines stand for continuous (second order) phase boundaries, while the dashed lines denote the first order phase boundaries. The black point represents the SU(2) symmetric isotropic Heisenberg limit. The $A1$ and $A3$ label Ising-like antiferromagnetic phases, with spins $S^z=\pm1/2$ and $S^z=\pm3/2$ aligned along the $z$ axis, while the $F3$ is the Ising-like ferromagnetic phase. The $SF_0$ is the easy plane N\'eel phase, the  $SF_A$ is a conical N\'eel phase, and $SF_F$ is a conical ferromagnetic phase.}
\label{fig:h0_pd}
\end{center}
\end{figure}

Let us start with the determination of the ground state phase diagram variationally, assuming a site factorized trial wave function $|\Psi\rangle=\prod_{i}|\Psi_i\rangle$, where the index $i$ runs over the lattice sites. The $|\Psi_i\rangle$ is a wave function in the four dimensional local Hilbert space of the $S=3/2$ spin on site $i$.\footnote{In the case of the $S=1/2$ spins, this approach is equivalent to treating the spin operators $S^\alpha$ classically ($\alpha = x,y,z$.}
 The variational phase diagram of the Hamiltonian (\ref{eq:Hamiltonian}) has been discussed previously in Ref.~\onlinecite{Romhanyi2011}: in accordance with experimental findings,\cite{Zheludev2003} a two-sublattice antiferromagnetic order is realized for the relevant parameters, with $|\Psi_i\rangle=|\Psi_A\rangle$ if site $i$ is on $A$ sublattice and $|\Psi_i\rangle=|\Psi_B\rangle$ for spins on $B$ sublattice. The phase diagram is shown in  Fig. \ref{fig:h0_pd}.

At positive values of $J_z$, the phases can be characterized by the staggered magnetic and superfluid order parameters: $m^{\text{st}}_{z}=\frac{1}{2}| S^{z}_A- S^{z}_B|$, and $O_{U(1)}=\frac{1}{2}|{\bf S}^{\bot}_A-{\bf S}^{\bot}_B|$, respectively, where ${\bf S}^{\bot}_j=(S^x_j,S^y_j)$. The partially and fully polarized axial (or Ising like) antiferromagnetic phases ($A1$ and $A3$, respectively) exhibit finite staggered magnetization. Additionally, superfluid phases appear that break the U(1) (or O(2)) symmetry: In the planar superfluid phase $SF_0$ ({\it i.e.} the easy-plane N\'eel ordered phase realized in the Ba$_2$CoGe$_2$O$_7$) only the fluid order parameter $O_{U(1)}$ is non-zero, while in the conical superfluid phase $SF_A$ both the $m^{\text{st}}_{z}$ and $O_{U(1)}$ have finite expectation values. In the conical ferromagnetic phase the finite magnetization along the $z$ direction coexist with a finite $O_{U(1)}$.
A first order transition line separates the planar superfluid phase $SF_0$ from the $A3$ and $F3$ gapped phases for smaller value of $\Lambda$, while for larger values of the on-site anisotropy, the $SF_0$ is bordered with the canted superfluid states via a second order transition line.

We remark that the $J_z \leftrightarrow -J_z$ symmetry of the phase diagram can be understood as follows: assuming coplanar spin structure in a plane $\Sigma$ perpendicular to the easy plane, rotating the spins on one of the sublattice by an angle $\pi$ around the axis perpendicular to the $\Sigma$, we change the sign of the $J_z$, the conical antiferromagnet SF$_{\text{A}}$ changes to a canted phase, while the $J$ and $\Lambda$ remain unchanged. Thus we can map the $J_z>0$ part of the phase diagram to the $J_z<0$ part, as seen in Fig.~\ref{fig:h0_pd}.

 Within the variational approach, the large $\Lambda$ stabilizes the easy-plane N\'eel-state ({\it i.e.},  $SF_0$ planar superfluid) ground state even for relatively large exchange anisotropy ($J_z/J\lessapprox 4$), with the following trial wave-function:
\begin{eqnarray} 
|\Psi_A\rangle &=& e^{-i \varphi_A \hat S^z} |\Psi_{\text{SF}}\rangle, 
\nonumber\\
|\Psi_B\rangle &=& e^{-i \varphi_B \hat S^z} |\Psi_{\text{SF}}\rangle, 
\label{eq:gs_AB}
\end{eqnarray}
where the $|\Psi_{\text{SF}}\rangle$ is characterized by the single variational parameter $\eta$,
\begin{equation}
|\Psi_{\text{SF}}\rangle = 
\frac{|\frac{3}{2}\rangle + \sqrt{3}\eta |\frac{1}{2}\rangle + \sqrt{3}\eta |-\frac{1}{2}\rangle + |-\frac{3}{2}\rangle}{\sqrt{6 \eta^2 + 2}}.
\label{eq:SF0_grst}
\end{equation}
The angles $\varphi_A$ and $\varphi_B$ measure the tilting of the spins from the $x$-axis and can be written as 
\begin{eqnarray}
  \varphi_A = \frac{\pi}{2}+ \phi \;, \quad
  \varphi_B = -\frac{\pi}{2}+ \phi  \;,
  \label{eq:phiAB}
\end{eqnarray}
so that $\varphi_A-\varphi_B=\pi$. This describes antiparallel spins (N\'eel-state) on the two sublattices, as the expectation value of the spin components on the A sublattice is 
\begin{equation}
\langle\Psi_A| \hat {\bf S} |\Psi_A\rangle = \frac{3 \eta (\eta +1)}{3 \eta^2+1} (\cos \varphi_A, \sin\varphi_A,0)\;,
\end{equation}
with an analogous expression for the spins on the B sublattice. For $\eta=1$, the $|\Psi_A\rangle$ and $|\Psi_B\rangle$ are spin-coherent states, with spin length 3/2. As the parameter $\eta$ increases, the supression of the $|\pm 3/2\rangle$ states reduces the length of the spin, and the wave function describes a spin of mixed dipolar, quadrupolar, and octupolar character. The independent parameter $\phi$ in Eq.~(\ref{eq:phiAB}) carries the O(2) symmetry breaking property of the superfluid phase (note that $\mathcal{H}$ commutes with $\hat S^z$) and it is the direction of the spins with respect to the $y$-axis. For convenience, we choose $\phi=0$ hereinafter.

The expectation value of the energy per site in the easy-plane N\'eel-state, as the function of parameter $\eta$, reads
\begin{eqnarray}
\frac{E^{SF_0}(\eta)}{N} = \frac{3}{4} \frac{\eta^2 + 3}{3 \eta^2 + 1} \Lambda -\frac{18 \eta^2
 \left( \eta + 1 \right)^2}{\left( 3\eta^2 + 1 \right)^2} J \;,
\label{eq:SF_en0}
 \end{eqnarray}
 where $N$ denotes the number of spins.
The $E^{SF_0}(\eta)$ becomes minimal when the condition 
\begin{equation}
\frac{\Lambda}{J} = \frac{3 (\eta^2 -1) (3 \eta +1)}{3 \eta^2+1}
\label{eq:eta_sol}
\end{equation}
is fulfilled. For only the in-plane spin components $S^x$ and $S^y$ are finite, the energy of the planar superfluid phase ($E^{SF_0}(\eta)$) does not depend on the $J_z$ term, which is reflected in the $\eta$ being dependent on the $\Lambda/J$ only. The usual procedure is to solve for the value of $\eta$ as a function of the parameters of the Hamiltonian --- in our case it would mean  finding the solution of a cubic polynomial. In the following we will rather use the parameter $\eta$ in the expressions instead of $\Lambda$ as given by Eq.~(\ref{eq:eta_sol}), and replace the relevant $\eta$ that corresponds to a specific $\Lambda$ value only at the end.

The first order transition between the antiferromagnetic $A3$ or ferromagnetic $F3$ Ising phases and the easy plane-N\'eel antiferromagnet  happens when $E^{{SF}_0}$ is equal to $E^{A3,F3}=9 \Lambda /4 \mp 9 J_z/2$ (upper sign corresponds to $A3$, lower to $F3$ Ising phase), conditions that provide the phase boundaries
\begin{equation}
J_z^{\text{1st}} = \pm J \frac{4 \eta^3 ( 3 \eta^2 + 2 \eta - 1)}{(3 \eta^2 + 1)^2} \;.
\end{equation} 

Let us briefly comment on the two limiting cases of the single-ion anisotropy.
(i) When $\Lambda/J\to 0$ the $\eta \to 1$, and the solution is the spin-coherent state mentioned earlier. Furthermore, for $J=J_z$ the Hamiltonian (\ref{eq:Hamiltonian}) recovers the O(3) symmetry and the ground state is the spin-3/2 N\'eel-state. 
(ii) In the limit $\Lambda/J\to \infty$ the variational parameter $\eta\to\infty$, the $S^z=\pm 3/2$ states of the spins are suppressed and the ground state wave function is composed of the $S^z=\pm 1/2$ states. Allowing for a general wave function in this reduced (two dimensional per site) Hilbert space, the tip of the spin spans a surface of an oblate ellipsoid. The length of the spin is maximal (equal to 1) when it lays in the $xy$-plane and minimal (equal to 1/2) along the $z$-axis, therefore a finite antiferromagnetic exchange selects ordering in the $xy$-plane.

\section{Multiboson spin-wave spectrum}\label{sec:flavor}

The usual spin-wave theory is a $1/S$ expansion in the length of the spin, where a single Holstein-Primakoff boson is introduced to describe the transversal fluctuations about the classical ground state. As we shall see shortly, the multiboson spin-wave approach supports the inclusion of more bosons that have essentially different nature, providing a powerful method to discuss higher order, i.e. quadrupole- or octupole-type excitations. 
\cite{ISI:A1985AXC5300038,Chubukov1990,N1988367}

Let us first introduce the bosons $\alpha^\dagger_m$ that create the $S^z=m$ states of the $S=3/2$ spin: $|m \rangle = \alpha^\dagger_{m} |\text{vacuum}\rangle$. Using the four $\alpha$ bosons, the diagonal spin operators can be written as
\begin{equation}
S^z=\sum_{m=-3/2}^{3/2} m \alpha^{\dagger}_{m} \alpha^{\phantom{\dagger}}_{m},
\quad
(S^z)^2 = \sum_{m=-3/2}^{3/2} m^2 \alpha^{\dagger}_{m} \alpha^{\phantom{\dagger}}_{m} .
\label{eq:s_a_diag}
\end{equation}
and the off-diagonal spin raising operator is
\begin{equation}
S^+ = 
 \sqrt{3} \left(\alpha^{\dagger}_{3/2} \alpha^{\phantom{\dagger}}_{1/2} + \alpha^{\dagger}_{-1/2} \alpha^{\phantom{\dagger}}_{-3/2} \right)
 +2 \alpha^{\dagger}_{1/2} \alpha^{\phantom{\dagger}}_{-1/2} ,
 \label{eq:s_a_offdiag}
\end{equation}
while the spin lowering operator can be easily obtained by its hermitian conjugate. All the spin operators (including higher order polynomials) can  be expressed as quadratic forms, so that they keep the number of bosons $M$ on each site conserved ($M=\sum_{m} \alpha^{\dagger}_{m,j} \alpha^{\phantom{\dagger}}_{m,j}$, and $M=1$ for the $S=3/2$ spin). As a consequence, the Hamiltonian (\ref{eq:Hamiltonian}) also commutes with the $\sum_{m} \alpha^{\dagger}_{m,j} \alpha^{\phantom{\dagger}}_{m,j}$.
Furthermore, written in this form, all the operators obey the expected  commutation relations.

For we want to carry out the multiboson spin-wave method starting from the planar antiferromagnetic ground state, we apply an SU(4) rotation in the space of $\alpha^\dagger_m$ bosons:
\begin{widetext}
\begin{eqnarray}
a^{\dagger}_j &=& \frac{1}{\sqrt{6 \eta^2+2}} \left[
e^{\frac{3}{2} i \varphi_j } \alpha^{\dagger}_{-3/2}
+e^{-\frac{3}{2} i \varphi_j} \alpha^{\dagger}_{3/2}
+\sqrt{3} \eta \left( e^{\frac{1}{2} i \varphi_j}  \alpha^{\dagger}_{-1/2}
+ e^{-\frac{1}{2}i \varphi_j }  \alpha^{\dagger}_{1/2} \right)
\right],
\label{eq:aboson}\\
b^{\dagger}_j &=& \frac{1}{\sqrt{14 \eta^2-8 \eta +2}} \left[
\sqrt{3} \eta \left( e^{\frac{3}{2} i \varphi_j } \alpha^{\dagger}_{-3/2}
  - e^{-\frac{3}{2} i \varphi_j } \alpha^{\dagger}_{3/2} \right)
+(2 \eta -1) \left( 
  e^{\frac{1}{2} i \varphi_j } \alpha^{\dagger}_{-1/2}
  -e^{-\frac{1}{2} i \varphi_j } \alpha^{\dagger}_{1/2} 
  \right)
\right],
\label{eq:bboson}\\
c^{\dagger}_j &=& 
\frac{1}{\sqrt{6 \eta^2+2}}
\left[
\sqrt{3} \eta \left(
  e^{\frac{3}{2} i \varphi_j } \alpha^{\dagger}_{-3/2} 
+ e^{-\frac{3}{2} i \varphi_j } \alpha^{\dagger}_{3/2}
\right)
- \left( e^{\frac{1}{2} i \varphi_j }\alpha^{\dagger}_{-1/2}
+ e^{-\frac{1}{2} i \varphi_j }\alpha^{\dagger}_{1/2} \right)
\right],
\label{eq:cboson}\\
d^{\dagger}_j &=& \frac{1}{\sqrt{14 \eta^2-8 \eta +2}} \left[
(2 \eta -1) \left( e^{\frac{3}{2} i \varphi_j } \alpha^{\dagger}_{-3/2}
-e^{-\frac{3}{2} i \varphi_j } \alpha^{\dagger}_{3/2}  \right)
-\sqrt{3} \eta \left(e^{\frac{1}{2} i \varphi_j } \alpha^{\dagger}_{-1/2}
 - e^{-\frac{1}{2} i \varphi_j }  \alpha^{\dagger}_{1/2} \right)
\right],
\label{eq:dboson}
\end{eqnarray}
\end{widetext}
In this rotated basis the variational ground state given by Eq.~(\ref{eq:gs_AB}) corresponds to the $|\Psi_A\rangle=a^\dagger_j|\text{vacuum}\rangle$ with $\varphi_j=\varphi_A = \pi/2$ for spins on $A$ sublattice and $|\Psi_B\rangle=a^\dagger_B|\text{vacuum}\rangle$ with $\varphi_j=\varphi_B = -\pi/2$. The $b$, $c$, and $d$ are suitably chosen bosons that will play the role of the Holstein-Primakoff bosons and describe the excitations of the system. Namely, inverting Eqs.~(\ref{eq:aboson})-(\ref{eq:dboson}) we can express the spin operators [Eqs.~(\ref{eq:s_a_diag}) and (\ref{eq:s_a_offdiag})]
using the $a$, $b$, $c$ and $d$ bosons, and replacing
\begin{eqnarray}
a^\dagger_j&\to&\sqrt{M-b^\dagger_j b^{\phantom{\dagger}}_j-c^\dagger_j c^{\phantom{\dagger}}_j-d^\dagger_j d^{\phantom{\dagger}}_j}
\label{eq:HPsuba}
\\
a^{\phantom{\dagger}}_j&\to&\sqrt{M-b^\dagger_j b^{\phantom{\dagger}}_j-c^\dagger_j c^{\phantom{\dagger}}_j-d^\dagger_j d^{\phantom{\dagger}}_j} 
\label{eq:HPsubb}
\end{eqnarray}
the spin operators (see Appendix~\ref{sec:appendix0}) still follow the expected commutation relations, analogously to  the familiar Holstein--Primakoff transformation that uses a single boson. Performing an expansion in the parameter $1/M$, one can further follow the procedure of the conventional spin wave theory for ordered magnets. The multiboson spin-wave Hamiltonian up to quadratic order in bosons then reads
\begin{eqnarray}
\mathcal{H} = M^2 \mathcal{H}^{(0)}+ M^{3/2} \mathcal{H}^{(1)}
 + M \mathcal{H}^{(2)} + O(M^{1/2}) \;,
\end{eqnarray}
where $\mathcal{H}^{(0)}$ is equal to the variational (or, equivalently, the mean field) energy Eq.~(\ref{eq:SF_en0}) and the $\mathcal{H}^{(1)}$ is identically zero when the variational condition~(\ref{eq:eta_sol}) is satisfied. The $\mathcal{H}^{(2)}$ is quadratic in bosonic operators and can be written as a sum of Hamiltonians in the $k$ space,
\begin{equation}
\mathcal{H}^{(2)} = \frac{1}{2} \sum_{{\bf k} \in {\rm BZ}} 
\left(\mathcal{H}^{(2)}_{bd,{\bf k}}+\mathcal{H}^{(2)}_{c,{\bf k}} 
\right)\;,
\end{equation}
where the $\mathcal{H}^{(2)}_{bd,{\bf k}}=\mathcal{H}^{(2)}_{bd,{\bf -k}}$ and $\mathcal{H}^{(2)}_{c,{\bf k}}=\mathcal{H}^{(2)}_{c,{\bf -k}}$ read
\begin{widetext}
\begin{eqnarray}
\mathcal{H}^{(2)}_{bd,{\bf k}} 
& = & 
\left[
\frac{6 (\eta+1)^2 \left(9 \eta ^3-5 \eta^2-\eta +1\right)}{\left(3\eta^2+1\right) \left(7\eta^2-4\eta+1\right)} J   
  -\frac{3 \left(7\eta^2-4\eta+1\right)}{3 \eta^2+1} J \gamma_{\bf k} 
 +\frac{12 \eta^2 (\eta+1)^2}{\left(3\eta^2+1\right) \left(7\eta^2-4\eta+1\right)} J_z \gamma_{\bf k} 
\right]
\left(
  b_{\bf k}^{\dagger} b_{\bf k}^{\phantom{\dagger}} 
+ b_{\bf -k}^{\dagger} b_{\bf -k}^{\phantom{\dagger}} 
\right)
\nonumber\\&&
+\left[\frac{12  \eta^2 (\eta+1)^2}{\left(3\eta^2+1\right) \left(7\eta^2-4\eta+1\right)} J_z +\frac{3 \left(7\eta^2-4\eta+1\right)}{3 \eta^2+1} J  \right] \gamma_{\bf k}
\left(
  b_{\bf k}^{\dagger} b_{\bf -k}^{\dagger} 
+ b_{\bf k}^{\phantom{\dagger}} b_{\bf -k}^{\phantom{\dagger}} 
 \right)
\nonumber\\&&
 +\frac{72 \eta ^3 (\eta+1)^2 }{\left(3\eta^2+1\right) \left(7\eta^2-4\eta+1\right)} J 
 \left(
   d_{\bf k}^{\dagger} d_{\bf k}^{\phantom{\dagger}}
 + d_{\bf -k}^{\dagger} d_{\bf -k}^{\phantom{\dagger}}
     \right) 
 \nonumber\\&&
+6 \sqrt{3} \frac{\eta (\eta+1) (\eta-1)^2}{\left(3\eta^2+1\right)\left(7\eta^2-4\eta+1\right)}
\left(
6 \eta   J -  J_z \gamma_{\bf k}
\right)
\left(
b_{\bf k}^{\dagger} d_{\bf k}^{\phantom{\dagger}} 
+ b_{\bf -k}^{\dagger} d_{\bf -k}^{\phantom{\dagger}} 
+ d_{\bf k}^{\dagger} b_{\bf k}^{\phantom{\dagger}} 
+ d_{\bf -k}^{\dagger} b_{\bf -k}^{\phantom{\dagger}} 
 \right),
\nonumber\\&&
+\frac{9 (\eta-1)^4}{\left(3\eta^2+1\right) \left(7\eta^2-4\eta+1\right)}  J_z \gamma_{\bf k}
 \left( 
 d_{\bf k}^{\dagger} d_{\bf -k}^{\dagger} 
+ d_{\bf k}^{\phantom{\dagger}} d_{\bf -k}^{\phantom{\dagger}} 
+ d_{\bf k}^{\dagger} d_{\bf k}^{\phantom{\dagger}} 
+ d_{\bf -k}^{\dagger} d_{\bf -k}^{\phantom{\dagger}} 
 \right)
\nonumber\\&&
-\frac{6 \sqrt{3} \eta (\eta+1)  (\eta-1)^2}{\left(3\eta^2+1\right) \left(7\eta^2-4\eta+1\right)} J_z \gamma_{\bf k}
\left(
 b_{\bf k}^{\phantom{\dagger}} d_{\bf -k}^{\phantom{\dagger}} 
+ b_{\bf -k}^{\phantom{\dagger}} d_{\bf k}^{\phantom{\dagger}} 
+ b_{\bf k}^{\dagger} d_{\bf -k}^{\dagger} 
+ b_{\bf -k}^{\dagger} d_{\bf k}^{\dagger} 
 \right),
 \label{eq:Hbd}\\
%
%
\mathcal{H}^{(2)}_{c,{\bf k}} 
& = & 
6 J (\eta+1) \left( 
c_{\bf k}^{\dagger} c_{\bf k}^{\phantom{\dagger}} + c_{\bf -k}^{\dagger} c_{\bf -k}^{\phantom{\dagger}} 
\right)
-\frac{3 (3\eta+1)^2 (\eta-1)^2 }{\left(3\eta^2+1\right)^2} J \gamma_{\bf k}
\left( 
c_{\bf k}^{\dagger} c_{\bf -k}^{\dagger} 
+ c_{\bf k}^{\phantom{\dagger}} c_{\bf -k}^{\phantom{\dagger}} 
+ c_{\bf k}^{\dagger} c_{\bf k}^{\phantom{\dagger}} 
+  c_{\bf -k}^{\dagger} c_{\bf -k}^{\phantom{\dagger}} 
\right). \label{eq:Hc}
\end{eqnarray}
\end{widetext}
In the above Hamiltonian we replaced $\Lambda$ by the expression for the $\eta$, Eq.~(\ref{eq:eta_sol}). The 
\begin{eqnarray} 
b^{\dagger}_{\bf k} &=& 
  \frac{1}{\sqrt{N}}\sum_{j}
     e^{-i{\bf k}\cdot {\bf r}_j}b^{\dagger}_{j} \;, \nonumber\\
b^{\phantom{\dagger}}_{\bf k} &=& 
  \frac{1}{\sqrt{N}}\sum_{j}
     e^{i{\bf k}\cdot {\bf r}_j}b^{\phantom{\dagger}}_{j} \;, 
\end{eqnarray}
is the $b^\dagger$ and $b$ bosonic operator in the momentum space, with analogous equations for the $c$ and $d$ bosons. Setting the lattice constant to 1, the geometrical factor $\gamma_{\bf k}$ can be expressed as
\begin{eqnarray}
\gamma_{\bf k}=\frac{1}{2} \left(\cos k_x +\cos k_y\right) \;.
\label{eq:gamma}
\end{eqnarray}
Note that $\gamma_{\bf k} = -\gamma_{\bf Q+k}$, where ${\bf Q} = (\pi,\pi)$ is the N\'eel ordering vector. Furthermore, for ${\bf k}\to 0$, $2 \sqrt{1-\gamma_{\bf k}} \to |{\bf k}| = k$, similarly $2 \sqrt{1+\gamma_{\bf Q+k}} \to |{\bf Q+ k}|$ as ${\bf k}\to {\bf Q}$.

\begin{figure}[tb]
\begin{center}
\includegraphics[width=8.5cm]{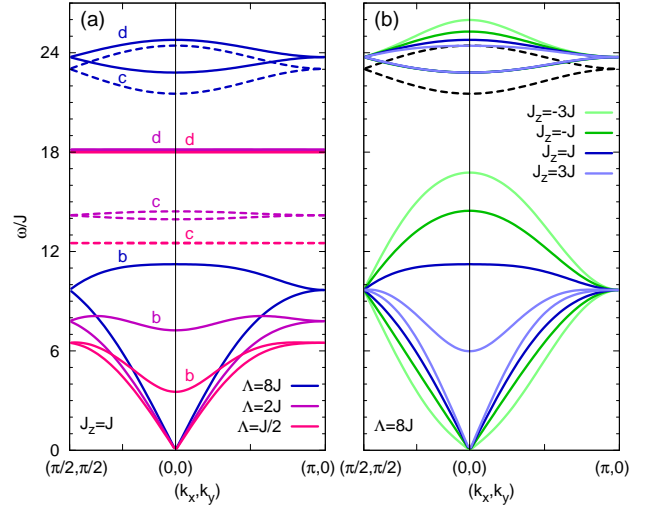}
\caption{(color online) Multiboson spin-wave dispersions in the reduced Brillouin-zone for (a) different single-ions anisotropies at $J=J_z$ and (b) different exchange anisotropies when $\Lambda = 8J$. The labels `b' and `d' denote the two branches that are eigenvalues of the $\mathcal{H}_{bd,{\bf k}}^{(2)}$ [Eq.~(\ref{eq:Hbd})]. The `c' (dashed lines) labels the eigenvalue of the $\mathcal{H}_{c,{\bf k}}^{(2)}$ [Eq.~(\ref{eq:Hc})] which is independent of the value of the exchange anisotropy $J_z$.
}
\label{fig:dipersion}
\end{center}
\end{figure}

The spin-wave Hamiltonians~(\ref{eq:Hbd}) and (\ref{eq:Hc}) can be diagonalized using Bogoliubov transformation. The excitation spectra obtained by numerical Bogoliubov transformation are shown in Fig.~\ref{fig:dipersion} for a set of selected $\Lambda$ and $J_z$ values.
For a given value of $\gamma_{\bf k}$ we get three eigenvalues (i.e., six modes for each ${\bf k}$ in the reduced Brillouin zone, as shown in the figures), one from $\mathcal{H}_c^{(2)}$ (we will denote this mode by letter `c') and two from $\mathcal{H}_{bd}^{(2)}$ (the `b' and `d' mode). This notation for the modes can also be traced back to the $\Lambda \to 0$ limit, where these modes stem from the $\eta \to 1$ form of the bosons $b$, $c$, and $d$ [Eqs.~(\ref{eq:bboson})-(\ref{eq:dboson})] (see Appendix \ref{sec:appendixA} for the discussion of the $\Lambda\to 0$ limit).  The `b' band is the lowest in energy and goes linearly to 0 at the ${\bf k}=(0,0)$ wave vector in the reduced Brillouin zone. 
The `c' and `d' bands are both gapped and higher in energy than the `b' band, typically `d' being the highest. Their dispersion is much smaller than that of the `b' band, and disappears as we decrease the single ion anisotropy $\Lambda$, becoming flat (localized) for $\Lambda=0$. 

 The eigenvalue of the $\mathcal{H}_c^{(2)}$ can be calculated analytically. It is independent of $J_z$ and reads
\begin{eqnarray}
\omega_{c}=6 J (\eta +1)
  \sqrt{1-\frac{(\eta-1)^2(3 \eta + 1)^2 }{(\eta +1)  \left(3 \eta ^2+1\right)^2}\gamma_{\bf k} } \;.
\end{eqnarray}
The analytical expression for the eigenvalues of $\mathcal{H}_{bd}^{(2)}$ is beyond our reach, except for two special cases: (i) along the lines $k_x+k_y= \pi$, when $\gamma_{\bf k}$ becomes zero and the energies are:
\begin{eqnarray}
\omega_{b,d}&=&
  \frac{3 J (\eta+1)^2 (3\eta+1)}{3\eta^2+1}
\nonumber\\&&
  \pm\frac{3 J (\eta+1) \sqrt{9\eta^4 - 24\eta^3 + 22\eta^2 + 8\eta + 1}}{3 \eta^2+1},
\end{eqnarray}
(ii) at ${\bf k} = (0,0)$ the $\gamma_{\bf k}=1$, and one of the eigenmodes is the $\omega_b=0$ Goldstone mode associated with turning the order parameter in the $xy$ plane and desribed by the self-adjoint operator $b^{\dagger}_{\text {GM}}=b^{\phantom{\dagger}}_{\text {GM}}$,
\begin{equation}
b^{\dagger}_{\text {GM}} \propto 
b_{(0,0)}^{\dagger} +b_{(0,0)}^{\phantom{\dagger}}  
-\frac{\sqrt{3} (\eta -1)^2}{2 \eta  (\eta +1) }
   \left( d_{(0,0)}^{\dagger} +d_{(0,0)}^{\phantom{\dagger}}   \right)\;,
\end{equation}
that commutes with the spin-wave Hamiltonian $\mathcal{H}^{(2)}$. 
At the same time, the `d' branch has energy
\begin{equation}
  \omega_d = 18 \frac{\eta+1}{3\eta^2+1} \sqrt{\eta (\eta^3-\eta^2+3\eta+1)} \;.
\end{equation}
In both of the above cases the energies are independent of the exchange anisotropy $J_z$. Apart from these special instances, analytical result are available in the $\Lambda\to 0$ and $\Lambda\to +\infty$ limits, as shown in Apendices \ref{sec:appendixA} and \ref{sec:appendixB}.

In Fig.~\ref{fig:dipersion}(b) we show the evolution of the dispersion as we change $J_z$, while we keep $\Lambda$ constant ($\Lambda=8J$ in the figure). It is the `b' band that is the most sensitive to value of $J_z$, while the higher energy `d' is only weakly affected. The energy of the `b' mode is linear in momentum, $\omega_b = v_b k$ as $k \to 0$, where the velocity can be calculated analytically by expanding the Hamiltonian in $\sqrt{1-\gamma_{\bf k}}$ and reads  
\begin{equation}
 v_b = \frac{6 \eta (\eta+1)}{3\eta^2+1}
  \sqrt{\frac{4 \eta^3}{\eta^3 - \eta^2 +3 \eta+ 1}J+J_z} \;. 
 \label{eq:vb}
\end{equation}
Increasing $J_z$, the energy of the `b' mode decreases at $\gamma_{\bf k}=-1$, and becomes 0 at a critical value $J_z = J_z^c$, where 
\begin{equation}
 J_z^c = J \frac{4 \eta^3}{\eta^3  - \eta^2 + 3 \eta + 1  }.
\end{equation}
The behavior of the `b' mode can be connected to the phase boundary of the easy-plane N\'eel phase: its softening marks the line of the second order transition into the conical N\'eel phase (or superfluid SF$_A$), as shown in the variational phase diagram, Fig.~\ref{fig:h0_pd}. 
Quite interestingly, for ferromagnetic $J_z$ Ising coupling, the second order phase transition line into the conical canted phase that is given by $J_z = - J_z^c$, is indicated by the vanishing of the spin wave velocity $v_b$, Eq.~(\ref{eq:vb}).

In order to test the reliability of the multiboson spin-wave method, we calculated the expectation number of bosons in the ground state, $\langle b^\dagger b^{\phantom{\dagger}} + c^\dagger c^{\phantom{\dagger}} + d^\dagger d^{\phantom{\dagger}}\rangle$. We found that the quantum fluctuations are the greatest and the boson expectation value are the largest in the fully isotropic ($\Lambda=0$, $J_z=J$) case: $\langle b^\dagger b^{\phantom{\dagger}} + c^\dagger c^{\phantom{\dagger}} + d^\dagger d^{\phantom{\dagger}}\rangle = 0.197$. Introducing even a small anisotropy reduced this value considerably. Our result for the isotropic point coincides with the known result for the spin reduction $\Delta S=0.197$ in the square lattice, \cite{Manousakis1991} as the bosons $c$ and $d$ decouple from the system, and the boson $b$ plays the role of the standard Holstein-Primakoff magnon.

\section{Spin structure factor}\label{sec:Sq}

The intensity at energy $\omega$ and momentum ${\bf k}$ in inelastic neutron measurement is determined by the magnetic cross section 
\begin{equation}
\frac{d^2 \sigma({\bf k},\omega)}{d\omega d\Omega} \propto \sum_{\mu\nu}
\left( \delta_{\mu\nu}-\frac{k_\mu k_\nu}{k^2}\right) 
S^{\mu\nu}({\bf k},\omega)
\end{equation}
at zero temperature ($\mu,\nu = x,y$,or $z$), where 
\begin{equation}
S^{\mu\nu}(\omega,{\bf k}) = 
\sum_{f} 
 \langle 0 | S^\mu_{\bf -k} | f \rangle \langle f | S^\nu_{\bf k} | 0 \rangle 
\delta(\omega - \omega_f)
\end{equation}
is the dynamical spin structure factor, $| 0 \rangle$ is the ground state, and the summation is over the excited states $f$. The matrix elements and energies in the $S^{\mu\nu}(\omega,{\bf k})$ can be calculated using the spin wave theory. The $1/M$ expansion for the spin operators, using Eqs.~(\ref{eq:s_a_diag}), (\ref{eq:s_a_offdiag}) and Eqs.~(\ref{eq:aboson})-(\ref{eq:dboson}), reads 
\begin{eqnarray}
S_j^x &=& 
\mp\frac{i \sqrt{3} \sqrt{7\eta^2-4\eta+1}}{2\sqrt{3\eta^2+1}}
\left(  b^{\dagger}_j - b^{\phantom{\dagger}}_j \right)\sqrt{M} 
,
\\
S_j^y &=& \pm M \frac{3 \eta  (\eta +1)}{3 \eta^2+1} 
\nonumber\\&&
 \pm\sqrt{M} \frac{\sqrt{3} (\eta -1) (3 \eta +1)}{2 \left(3 \eta^2+1\right)}  \left( c^{\dagger}_j + c^{\phantom{\dagger}}_j \right),
\\
S_j^z &=& \sqrt{M} 
\left[
-\frac{\sqrt{3} \eta (\eta +1) }{\sqrt{3 \eta^2+1} \sqrt{7 \eta^2-4 \eta +1}} \left( b^{\dagger}_j + b^{\phantom{\dagger}}_j \right) \right.
\nonumber\\
&& \left.+\frac{3  (\eta -1)^2}{2 \sqrt{3 \eta^2+1} \sqrt{7 \eta^2-4 \eta +1}} \left( d^{\dagger}_j + d^{\phantom{\dagger}}_j\right)
\right],
\end{eqnarray}
where only the leading order terms that are proportional to $M$ and $\sqrt{M}$ are shown, and the upper (lower) sign corresponds to the spin on sublattice A (B).
Due to the alternating sign in $S_j^x$ and $S_j^y$, in the Fourier transform of these operators the Holstein-Primakoff bosons are shifted by the N\'eel-ordering vector  ${\bf Q}=(\pi,\pi)$: 
\begin{eqnarray}
  S^x_{\bf k} &\propto& -i\left(  b^{\dagger}_{{\bf k}+{\bf Q}} - b^{\phantom{\dagger}}_{-{\bf k}-{\bf Q}} \right)\sqrt{M} \\
  S^y_{\bf k} &\propto&  \left(  c^{\dagger}_{{\bf k}+{\bf Q}} - c^{\phantom{\dagger}}_{-{\bf k}+{\bf Q}} \right)\sqrt{M} \;,
\end{eqnarray}
correspondingly
\begin{equation}
S^{\mu\mu}(\omega,{\bf k}) =  \sum_{f} 
 \left| \langle f | S^\mu_{\bf k} | 0 \rangle \right|^2
\delta(\omega - \omega_{\bf k+Q}) 
\end{equation}
for $\mu = x,y$.
Such a momentum shift is not needed for the $S^z_{\bf k}$. 

\begin{figure}[tb]
\begin{center}
\includegraphics[width=8cm]{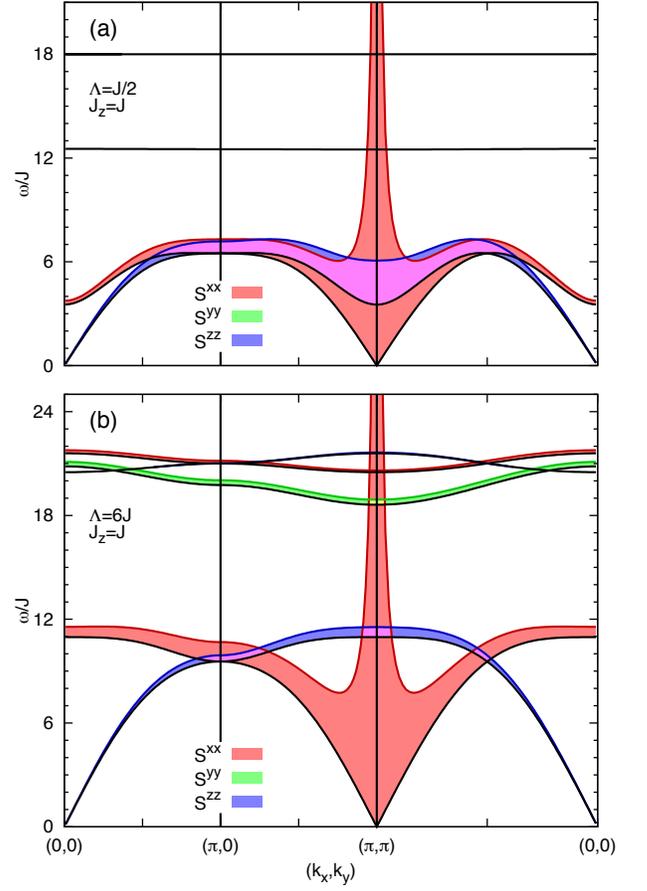}
\caption{(color online) Dynamical structure factor $S^{\mu\mu}({\bf k},\omega)$ with $\mu\in\left\{x,y,z\right\}$ for $J_z/J=1$ and (a) $\Lambda=J/2$ and (b) $\Lambda=6J$ along a path in the Brillouin zone. The widths of the filled curves above the excitation energies (black solid lines) denote the strength of the matrix elements $|\langle f| S^{\mu\mu}_{\bf k}|0\rangle|^2$. The $S^{xx}({\bf k},\omega)$ diverges as $1/\omega$ as ${\bf k\to Q}$. We note that the ground state is a N\'eel antiferromagnet with the spins chosen to be parallel to the $y$-axis. Therefore, the `c' is a stretching (or longitudinal) mode, associated with length fluctuations of the spins, while the `b' and `d' are transverse modes.}
\label{fig:intensity}
\end{center}
\end{figure}

\begin{figure}[tb]
\begin{center}
\includegraphics[width=8.5cm]{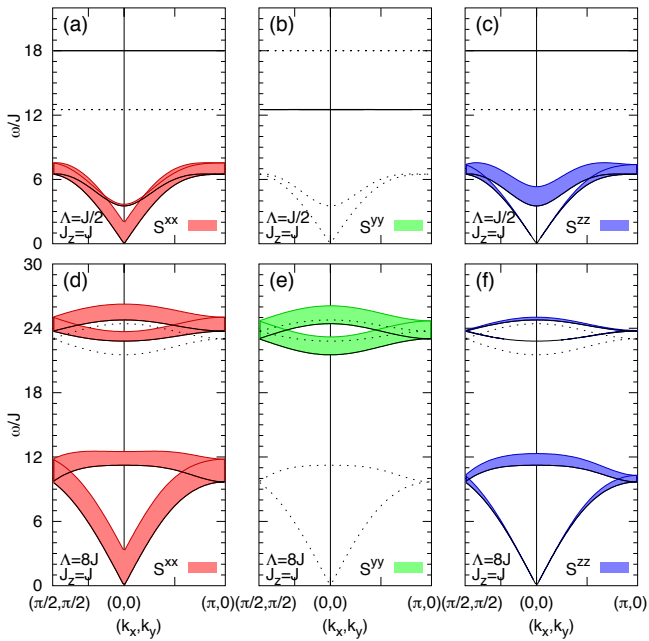}
\caption{(color online) The different components of the dynamical structure factor  are shown separately in the reduced Brillouin zone for the anisotropy parameter $\Lambda=J/2$ in (a)--(c) and $\Lambda=8J$ in (d)--(f), while in all cases $J_z/J=1$. Here the widths of the filled curves above the excitation energies denote the matrix elements $|\langle f| S^{\mu\mu}_{\bf k}|0\rangle|^2$ multiplied with energy $\omega_f$, {\it i.e.}, $\omega S^{\mu\mu}({\bf k},\omega)$.  The ground state is a N\'eel antiferromagnet with the spins chosen to be parallel to the $y$-axis. The dotted lines denote silent modes in the corresponding $S^{\mu\mu}({\bf k},\omega)$. }
\label{fig:intensityRB}
\end{center}
\end{figure}

The matrix elements of the different modes in the Brillouin zone of the square lattice, evaluated numerically, are shown in Fig.~{\ref{fig:intensity}} for a small and a large value of the single-ion anisotropy $\Lambda$. It turns out that the `b' and `d' modes have finite matrix elements with the $S^x$ and $S^z$ spin components that are perpendicular to the orientation of the spins, i.e. these modes are transversal modes, similarly to the modes in the conventional spin-wave theory. The `c' mode is more interesting, since the only nonvanishing matrix element is with the $S^y$ spin operator: in this mode the length of the spin changes ("spin stretching mode"). In all cases the $S^{xx}({\bf k},\omega)$ diverges as $1/\omega$ when ${\bf k}$ approaches the ${\bf Q}=(\pi,\pi)$ ordering wave vector --- this reflects the zero energy cost of rotating the spins in the easy plane, {\it i.e.} the Goldstone-mode. On the other hand, the $S^{zz}({\bf k},\omega)$ associated with spin fluctuations perpendicular to the easy plane, while finite at ${\bf Q}$ for finite values of $\Lambda$, diverges as the anisotropy gap is closed. Eventually,  $S^{xx}$ and $S^{zz}$ become equal for $\Lambda=0$, when the full O(3) symmetry is recovered. Furthermore, as $\Lambda$ is decreased, the intensity of the `c' and `d' modes decreases rapidly (see Appendix~\ref{sec:appendixA} for a detailed discussion of the $\Lambda \to 0$ limit). To eliminate the $1/\omega$ divergence and obtain a better comparison of the matrix elements, we show $\omega S^{\mu\mu}({\bf k},\omega)$ in Fig.~(\ref{fig:intensityRB}), this time along a path in the reduced Brillouin zone.

\section{Comparison with B\lowercase{a}$_2$C\lowercase{o}G\lowercase{e}$_2$O$_7$ neutron scattering experiments}\label{sec:INS}

\begin{figure}[tb]
\begin{center}
\includegraphics[width=7cm]{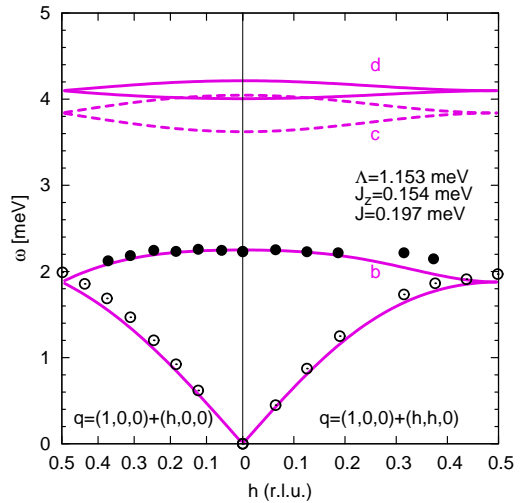}
\caption{(color online) The spin-wave dispersion for $\Lambda=1.15$ meV, $J_z=0.154$ meV, and $J = 0.197$ meV.
The wave vectors $q$ used in Ref.~\onlinecite{Zheludev2003} and in this figure for easier comparison are defined using the two-Co ion unit cell, so that ${\bf q} = (k_x+k_y,k_x-k_y,0)/2\pi$.
}
\label{fig:neutron_ba2coge2o7}
\end{center}
\end{figure}

Inelastic neutron scattering measurements on Ba$_2$CoGe$_2$O$_7$ were reported in Ref.~\onlinecite{Zheludev2003}. In the experiment constant-momentum scans were performed up to 3-4 meV (depending on momentum), and the peaks in the intensity were traced to get the dispersion. A single mode was observed with a dispersion of 2.2 meV and with a large anisotropy gap comparable to the dispersion itself. The mode was fitted using `conventional' spin-wave theory based on the Heisenberg model extended with strong exchange anisotropy only (i.e. $\Lambda=0$). No higher modes were observed in the aforementioned energy window.

On the other hand, additional peaks have been observed in far infrared spectra\cite{Kezsmarki2011} at around 1THz beside the 0.5 THz mode that corresponds to the 2.2 meV peak detected by inelastic neutron scattering (we use that 1Thz $\approx$ 4.13meV). Furthermore, a recent study of the far infrared absorption in high magnetic field with high resolution revealed several modes that could be described by the Hamiltonian~(\ref{eq:Hamiltonian}) using the presented multi boson model \cite{FIR2012}. Therefore we anticipate that those lines shall also be present in the inelastic neutron spectra. To be more precise, in Fig.~\ref{fig:neutron_ba2coge2o7} we compare the calculated multiboson dispersions using the parameters $\Lambda=1.15$ meV, $J_z=0.154$ meV, and $J = 0.197$ meV of Ref.~\onlinecite{FIR2012} with the inelastic neutron scattering  peaks taken from Ref.~\onlinecite{Zheludev2003}. As shown, the ${\bf k}$ dependence is nicely reproduced, and we expect the additional, weakly dispersing peaks with smaller intensity at energies that are about 4 meV.

\section{Conclusions}\label{sec:conclusions}

In this paper we have extended the multiboson spin-wave theory to the spin-3/2 N\'eel antiferromagnet with strong single-ion anisotropy: the spin operators in the ordered state are described by three Holstein-Primakoff bosons, and accordingly,  the excitation spectrum in this approach consist of three modes for each spin. 

In the absence of the single-ion anisotropy, only one mode out of the three has finite matrix elements with the spin-dipole operators. This mode is equivalent to the magnon mode of the `conventional' spin-wave theory in every aspect. The additional two modes, silent in spectroscopy probes that interact with the magnetic moment only (such as neutron scattering), describe quadrupolar and octupolar fluctuations of the $S=3/2$ spin and may become visible when these multipolar fluctuations couple, for example, to electric polarization. This may happen if the spins are in a non-centrosymmetric environment, where coupling between the quadrupolar spin fluctuation and electric polarization makes them visible in light absorption experiments, such as in far infrared spectra \cite{Kezsmarki2011,FIR2012}. 

The picture above changes markedly when finite single-ion anisotropy is present in the system. As a result of the single-ion anisotropy, the spins in the mean field (equivalently site--factorized variational wave function) approximation are not any more spin-coherent states in the N\'eel ordered phase: the suppression of the $S^z=\pm 3/2$ spin-states for $\Lambda>0$ leads to shortening of the spins. This allows longitudinal fluctuations of the spins (stretching modes) that have finite matrix elements with the spin-dipole operator parallel to the ordered moment, thus become observable in neutron, electron spin resonance and other spectroscopies. 
Our findings are closely related to the observed longitudinal excitation mode in the pressure-induced in-plane AFM phase of NiCl$_2$-4SC(NH$_2$)$_2$.\cite{Matsumoto2007}
 A similar approach has been introduced to describe the nature of the excitations in the gapped TlCuCl$_3$ spin-dimer compound, where the one-magnon Raman scattering has been found efficient for selectively observing such longitudinal excitations in the pressure-induced ordered phases.\cite{Kuroe2008,Matsumoto2008}

Finally, we compared the multiboson spin-wave modes with the inelastic neutron spectra of Ba$_2$CoGe$_2$O$_7$. We found that the lowest mode  reproduces the measured dispersion using the exchange and on-site anisotropy parameters fitted from the evaluation of the far-infrared absorption measurements.\cite{FIR2012} We propose that the other, higher energy modes shall also be observed in neutron scattering experiments at  $\approx$4meV, with weak dispersions. From the analysis of the spin-structure factor the stretching modes could, in principle, be identified. Furthermore, we believe that such stretching modes shall appear in any $S>1/2$ material with strong single-ion anisotropy, starting from the related compounds Ca$_x$Sr$_{2-x}$CoSi$_2$O$_7$
\cite{Akai2009,Akai2010} and Ba$_2$MnGe$_2$O$_7$\footnote{H. Murakawa, private communication}.

\begin{acknowledgments}
We are pleased to thank R. Coldea, I. K\'ezsm\'arki, B. Lake, N. Shannon, F. Titusz, and M. Zhitomirsky for stimulating discussions. This work has been supported by Hungarian OTKA Grant Nos. K73455.
\end{acknowledgments}

\appendix

\section{Expansion of the spin operators}
\label{sec:appendix0}

Considering the $b_j$, $c_j$, and $d_j$ as Holstein-Primakoff bosons and replacing the $a^\dagger$ and $a$ bosons with the expressions (\ref{eq:HPsuba}) and (\ref{eq:HPsubb}), the spin operators on the A sublattice read
\begin{widetext}
\begin{eqnarray}
S^x_j&=&
\frac{i \sqrt{3} \sqrt{7 \eta ^2-4 \eta +1} } {2 \sqrt{3 \eta ^2+1}} 
\left(\sqrt{M - b^{\dagger}_j b^{\phantom{\dagger}}_j - c^{\dagger}_j c^{\phantom{\dagger}}_j - d^{\dagger}_j d^{\phantom{\dagger}}_j} b^{\phantom{\dagger}}_j- b^{\dagger}_j \sqrt{M - b^{\dagger}_j b^{\phantom{\dagger}}_j - c^{\dagger}_j c^{\phantom{\dagger}}_j - d^{\dagger}_j d^{\phantom{\dagger}}_j} \right)
\nonumber\\&&
+\frac{i \left(3 \eta ^2+2 \eta -1\right) }{\sqrt{3 \eta ^2+1} \sqrt{7 \eta ^2-4 \eta +1}} 
\left(b^{\dagger}_j c^{\phantom{\dagger}}_j - c^{\dagger}_j b^{\phantom{\dagger}}_j \right)
+\frac{i \sqrt{9 \eta ^2+3} }{2 \sqrt{7 \eta ^2-4 \eta +1}} 
\left(c^{\dagger}_j d^{\phantom{\dagger}}_j - d^{\dagger}_j c^{\phantom{\dagger}}_j \right) ,
\end{eqnarray}
\begin{eqnarray}
S^y_j&=&
+\frac{3 \eta (\eta +1) }{3 \eta ^2+1} 
\left(M - b^{\dagger}_j b^{\phantom{\dagger}}_j - c^{\dagger}_j c^{\phantom{\dagger}}_j - d^{\dagger}_j d^{\phantom{\dagger}}_j \right)
\nonumber\\&&
+\frac{\sqrt{3} \left(3 \eta ^2-2 \eta -1\right) }{6 \eta ^2+2} 
\left(\sqrt{M - b^{\dagger}_j b^{\phantom{\dagger}}_j - c^{\dagger}_j c^{\phantom{\dagger}}_j - d^{\dagger}_j d^{\phantom{\dagger}}_j} \, c^{\phantom{\dagger}}_j + c^{\dagger}_j \sqrt{M - b^{\dagger}_j b^{\phantom{\dagger}}_j - c^{\dagger}_j c^{\phantom{\dagger}}_j - d^{\dagger}_j d^{\phantom{\dagger}}_j} \right)
\nonumber\\&&
+\frac{\left(2 \eta ^2+\eta -1\right) }{7 \eta ^2-4 \eta +1} b^{\dagger}_j b^{\phantom{\dagger}}_j
+\frac{\sqrt{3} \left(5 \eta ^2-6 \eta +1\right) }{14 \eta ^2-8 \eta +2} 
\left(b^{\dagger}_j d^{\phantom{\dagger}}_j + d^{\dagger}_j b^{\phantom{\dagger}}_j\right)
+\frac{(1-3 \eta ) }{3 \eta ^2+1} c^{\dagger}_j c^{\phantom{\dagger}}_j
+\frac{3 (1-3 \eta ) \eta }{7 \eta ^2-4 \eta +1} d^{\dagger}_j d^{\phantom{\dagger}}_j ,
\end{eqnarray}
\begin{eqnarray}
S^z_j&=&
+\frac{3 (\eta -1)^2 }{2 \sqrt{3 \eta ^2+1} \sqrt{7 \eta ^2-4 \eta +1}} 
\left(\sqrt{M - b^{\dagger}_j b^{\phantom{\dagger}}_j - c^{\dagger}_j c^{\phantom{\dagger}}_j - d^{\dagger}_j d^{\phantom{\dagger}}_j} d^{\phantom{\dagger}}_j + d^{\dagger}_j \sqrt{M - b^{\dagger}_j b^{\phantom{\dagger}}_j - c^{\dagger}_j c^{\phantom{\dagger}}_j - d^{\dagger}_j d^{\phantom{\dagger}}_j} \right)
\nonumber\\&&
-\frac{\sqrt{3} \eta (\eta +1) }{\sqrt{3 \eta ^2+1} \sqrt{7 \eta ^2-4 \eta +1}} 
\left(\sqrt{M - b^{\dagger}_j b^{\phantom{\dagger}}_j - c^{\dagger}_j c^{\phantom{\dagger}}_j - d^{\dagger}_j d^{\phantom{\dagger}}_j} b^{\phantom{\dagger}}_j + b^{\dagger}_j \sqrt{M - b^{\dagger}_j b^{\phantom{\dagger}}_j - c^{\dagger}_j c^{\phantom{\dagger}}_j - d^{\dagger}_j d^{\phantom{\dagger}}_j} \right)
\nonumber\\&&
+\frac{\left(-9 \eta ^2+2 \eta -1\right) }{2 \sqrt{3 \eta ^2+1} \sqrt{7 \eta ^2-4 \eta +1}} 
\left(b^{\dagger}_j c^{\phantom{\dagger}}_j + c^{\dagger}_j b^{\phantom{\dagger}}_j \right)
-\frac{\sqrt{3} \eta (3 \eta -1) }{\sqrt{3 \eta ^2+1} \sqrt{7 \eta ^2-4 \eta +1}} 
\left(c^{\dagger}_j d^{\phantom{\dagger}}_j + d^{\dagger}_j c^{\phantom{\dagger}}_j \right),
\end{eqnarray}
and
\begin{eqnarray}
(S^z_j)^2&=&
+\frac{3 \left(\eta ^2+3\right)}{12 \eta ^2+4}
\left(M - b^{\dagger}_j b^{\phantom{\dagger}}_j - c^{\dagger}_j c^{\phantom{\dagger}}_j - d^{\dagger}_j d^{\phantom{\dagger}}_j \right)
\nonumber\\&&
+\frac{2 \sqrt{3} \eta }{3 \eta ^2+1} 
\left( \sqrt{M - b^{\dagger}_j b^{\phantom{\dagger}}_j - c^{\dagger}_j c^{\phantom{\dagger}}_j - d^{\dagger}_j d^{\phantom{\dagger}}_j} c^{\phantom{\dagger}}_j + c^{\dagger}_j \sqrt{M - b^{\dagger}_j b^{\phantom{\dagger}}_j - c^{\dagger}_j c^{\phantom{\dagger}}_j - d^{\dagger}_j d^{\phantom{\dagger}}_j} \right)
\nonumber\\&&
+\frac{\left(31 \eta ^2-4 \eta +1\right) }{28 \eta ^2-16 \eta +4} b^{\dagger}_j b^{\phantom{\dagger}}_j 
+\frac{2 \sqrt{3} \eta (2 \eta -1) }{7 \eta ^2-4 \eta +1} 
\left( b^{\dagger}_j d^{\phantom{\dagger}}_j + d^{\dagger}_j b^{\phantom{\dagger}}_j \right)
+\frac{\left(27 \eta ^2+1\right) }{12 \eta ^2+4} c^{\dagger}_j c^{\phantom{\dagger}}_j
+\frac{\left(39 \eta ^2-36 \eta +9\right) }{28 \eta ^2-16 \eta +4} d^{\dagger}_j d^{\phantom{\dagger}}_j .
\end{eqnarray}
As written, the expression above satisfy the exact commutation relation of the spin operators, i.e. no approximation has been made yet. 

The spin operators expanded in $1/M$, neglecting the $O(M^{-1/2})$ terms, are:
\begin{eqnarray}
S^x_j&=&
\frac{i \sqrt{3} \sqrt{7 \eta ^2-4 \eta +1} } {2 \sqrt{3 \eta ^2+1}} 
\sqrt{M} \left( b^{\phantom{\dagger}}_j- b^{\dagger}_j \right)
\nonumber\\&&
+\frac{i \left(3 \eta ^2+2 \eta -1\right) }{\sqrt{3 \eta ^2+1} \sqrt{7 \eta ^2-4 \eta +1}} 
\left(b^{\dagger}_j c^{\phantom{\dagger}}_j - c^{\dagger}_j b^{\phantom{\dagger}}_j \right)
+\frac{i \sqrt{9 \eta ^2+3} }{2 \sqrt{7 \eta ^2-4 \eta +1}} 
\left(c^{\dagger}_j d^{\phantom{\dagger}}_j - d^{\dagger}_j c^{\phantom{\dagger}}_j \right), 
\\
S^y_j&=&
\frac{3 \eta (\eta +1) }{3 \eta ^2+1} M
+\frac{\sqrt{3} \left(3 \eta ^2-2 \eta -1\right)}{6 \eta ^2+2} 
\sqrt{M} \left(c^{\phantom{\dagger}}_j + c^{\dagger}_j \right)
-\frac{3 \eta (\eta +1) }{3 \eta ^2+1} 
\left( b^{\dagger}_j b^{\phantom{\dagger}}_j + c^{\dagger}_j c^{\phantom{\dagger}}_j + d^{\dagger}_j d^{\phantom{\dagger}}_j \right)
\nonumber\\&&
+\frac{\left(2 \eta ^2+\eta -1\right) }{7 \eta ^2-4 \eta +1} b^{\dagger}_j b^{\phantom{\dagger}}_j
+\frac{\sqrt{3} \left(5 \eta ^2-6 \eta +1\right) }{14 \eta ^2-8 \eta +2} 
\left(b^{\dagger}_j d^{\phantom{\dagger}}_j + d^{\dagger}_j b^{\phantom{\dagger}}_j\right)
+\frac{(1-3 \eta ) }{3 \eta ^2+1} c^{\dagger}_j c^{\phantom{\dagger}}_j
+\frac{3 (1-3 \eta ) \eta }{7 \eta ^2-4 \eta +1} d^{\dagger}_j d^{\phantom{\dagger}}_j,
\\
S^z_j&=&
\frac{3 (\eta -1)^2 }{2 \sqrt{3 \eta ^2+1} \sqrt{7 \eta ^2-4 \eta +1}} 
\sqrt{M} \left(d^{\phantom{\dagger}}_j + d^{\dagger}_j \right)
-\frac{\sqrt{3} \eta (\eta +1) }{\sqrt{3 \eta ^2+1} \sqrt{7 \eta ^2-4 \eta +1}} 
\sqrt{M} \left( b^{\phantom{\dagger}}_j + b^{\dagger}_j \right)
\nonumber\\&&
+\frac{\left(-9 \eta ^2+2 \eta -1\right) }{2 \sqrt{3 \eta ^2+1} \sqrt{7 \eta ^2-4 \eta +1}} 
\left(b^{\dagger}_j c^{\phantom{\dagger}}_j + c^{\dagger}_j b^{\phantom{\dagger}}_j \right)
-\frac{\sqrt{3} \eta (3 \eta -1) }{\sqrt{3 \eta ^2+1} \sqrt{7 \eta ^2-4 \eta +1}} 
\left(c^{\dagger}_j d^{\phantom{\dagger}}_j + d^{\dagger}_j c^{\phantom{\dagger}}_j \right),
\\ 
(S^z_j)^2&=&
\frac{3 \left(\eta ^2+3\right)}{12 \eta ^2+4} M
+\frac{2 \sqrt{3} \eta }{3 \eta ^2+1} 
\sqrt{M} \left( c^{\phantom{\dagger}}_j + c^{\dagger}_j \right)
-\frac{3 \left(\eta ^2+3\right)}{12 \eta ^2+4}
\left(b^{\dagger}_j b^{\phantom{\dagger}}_j + c^{\dagger}_j c^{\phantom{\dagger}}_j + d^{\dagger}_j d^{\phantom{\dagger}}_j \right)
\nonumber\\&&
+\frac{\left(31 \eta ^2-4 \eta +1\right) }{28 \eta ^2-16 \eta +4} b^{\dagger}_j b^{\phantom{\dagger}}_j 
+\frac{2 \sqrt{3} \eta (2 \eta -1) }{7 \eta ^2-4 \eta +1} 
\left( b^{\dagger}_j d^{\phantom{\dagger}}_j + d^{\dagger}_j b^{\phantom{\dagger}}_j \right)
+\frac{\left(27 \eta ^2+1\right) }{12 \eta ^2+4} c^{\dagger}_j c^{\phantom{\dagger}}_j
+\frac{\left(39 \eta ^2-36 \eta +9\right) }{28 \eta ^2-16 \eta +4} d^{\dagger}_j d^{\phantom{\dagger}}_j .
\end{eqnarray}
\end{widetext}

\section{The case of small single-ion anisotropy}
\label{sec:appendixA}

When $\Lambda \ll J,J_z$, the parameter $\eta$ in Eq.~(\ref{eq:eta_sol}) that minimizes the energy can be expanded as
\begin{equation}
 \eta = 
1+\frac{\Lambda }{6 J}+\frac{\Lambda^2}{144 J^2}+O\left(\frac{\Lambda^3}{J^3}\right)  \;.
\end{equation}
Substituting this expression into the multiboson spin-wave Hamiltonian (\ref{eq:Hbd0}), we get $O(\Lambda^2)$ 
\begin{eqnarray}
\mathcal{H}^{(2)}_{bd,{\bf k}} 
& = & 
\left[
6 J   
  - 3\left(J - J_z \right)  \gamma_{\bf k} 
\right]
\left(
  b_{\bf k}^{\dagger} b_{\bf k}^{\phantom{\dagger}} 
+ b_{\bf -k}^{\dagger} b_{\bf -k}^{\phantom{\dagger}} 
\right)
\nonumber\\&&
+ 3\left(J  + J_z\right) 
\left(
  b_{\bf k}^{\dagger} b_{\bf -k}^{\dagger} 
+ b_{\bf k}^{\phantom{\dagger}} b_{\bf -k}^{\phantom{\dagger}} 
 \right)
\nonumber\\&&
   + \frac{\Lambda}{2} \left(2 
    -  \gamma_{\bf k}  - \frac{J_z}{J}  \gamma_{\bf k}  \right)
\left(
  b_{\bf k}^{\dagger} b_{\bf k}^{\phantom{\dagger}} 
+ b_{\bf -k}^{\dagger} b_{\bf -k}^{\phantom{\dagger}} 
\right)
\nonumber\\&&
+ \frac{\Lambda}{2} \left(1-\frac{J_z}{J} \right) \gamma_{\bf k}
\left(
  b_{\bf k}^{\dagger} b_{\bf -k}^{\dagger} 
+ b_{\bf k}^{\phantom{\dagger}} b_{\bf -k}^{\phantom{\dagger}} 
 \right)
\nonumber\\&&
 +18 J 
 \left(
   d_{\bf k}^{\dagger} d_{\bf k}^{\phantom{\dagger}}
 + d_{\bf -k}^{\dagger} d_{\bf -k}^{\phantom{\dagger}}
 \right),
 \label{eq:Hbd0}\\
\mathcal{H}^{(2)}_{c,{\bf k}} 
& = & 
\left( 12 J + \Lambda \right)
\left( 
c_{\bf k}^{\dagger} c_{\bf k}^{\phantom{\dagger}} + c_{\bf -k}^{\dagger} c_{\bf -k}^{\phantom{\dagger}} 
\right)\;.
 \label{eq:Hc0}
\end{eqnarray}
The $b$, $c$, and $d$ bosons decouple in this limit. The $b$ bosons propagate, while the $c$ and $d$ bosons are localized. The energy of the different modes (here we keep higher order terms in Eq.~(\ref{eq:Hbd0})) are
\begin{eqnarray}
 \ \omega^{b}_{\bf k} &=& 6 J \sqrt{1-\gamma_{\bf k}} \nonumber\\
 && \times \sqrt{1 + \gamma_{\bf k} \frac{J_z}{J} + \frac{\Lambda}{3J} + (J - J_z \gamma_{\bf k}) \frac{\Lambda^2}{72 J^3} + \dots }, \nonumber\\
 &&\\
 \omega^{c}_{\bf k} &=& 12 J + \Lambda + (1-2  \gamma_{\bf k})\frac{\Lambda^2}{24J} + \dots, \\
 \omega^{d}_{\bf k} &=& 18 J +  \frac{\Lambda^3}{48 J^2} +\dots , 
\end{eqnarray}
The $\omega_{b_+}$ shows the typical square-root behaviour of the anisotropy gap on the exchange anisotropy $J-J_z$ and single-ion anisotropy $\Lambda$. The $c$ and $d$ bosons are related to higher, $\Delta S^y=\pm 2$ and $\Delta S^y=\pm 3$ transitions in systems (quadrupolar and octupolar), as for $\eta=1$ the 
\begin{equation}
  S^y_j = \pm \frac{3}{2} M \mp \left(  b^{\dagger}_j b^{\phantom{\dagger}}_j + 2 c^{\dagger}_j c^{\phantom{\dagger}}_j  + 3 d^{\dagger}_j d^{\phantom{\dagger}}_j \right) .
\end{equation}
For $\Lambda=0$ there are only bilinear exchange terms in the Hamiltonian, therefore the quadrupolar and octupolar moments of the spin on neighboring sites do not interact.
The energy of the $\Delta S^y=\pm 2$ `c' mode is simply $12 J$, and that is actually the Zeeman energy that corresponds to this transition in the mean field of the four neighboring antialligned spin, each contributing by $(3/2) J$ to the Weiss field $h_{\text{Weiss}}=4 \times 3J/2 = 6J$. Correspondingly, the energy of the $d$-branch is $18 J = 3 h_{\text{Weiss}}$, the Zeeman enegy of the $\Delta S^y=\pm 3$ transition. Both `c' and `d' modes acquire dispersion in higher order in $\Lambda$: the mode `c' in the order $\Lambda^2/J$ and `d' in the order $\Lambda^4/J^3$. 

The matrix elements in the dynamical spin structure factor are given in this limit as 
\begin{eqnarray}
 |\langle 0 | S^x_{\bf k} | b \rangle |^2 &=& \frac{3}{4} \sqrt{\frac{J - J_z \gamma_k+\Lambda/3}{J+J \gamma_k}} + O(\Lambda^2) \\
 |\langle 0 | S^z_{\bf k} | b \rangle |^2 &=& \frac{3}{4} \sqrt{\frac{J-J \gamma_k}{J + J_z \gamma_k+\Lambda/3}} + O(\Lambda^2) \\
 |\langle 0 | S^y_{\bf k} | c \rangle |^2 &=& \frac{\Lambda^2}{48 J^2} + O(\Lambda^3)
\end{eqnarray}
while both $|\langle 0 | S^x_{\bf k} | d \rangle |^2$ and $|\langle 0 | S^z | d \rangle |^2$ are $O(\Lambda^4)$. 
The $|\langle 0 | S^x_{\bf k} | b \rangle |^2$ matrix element of the `b' mode, associated with the Goldstone mode, diverges as $1/\omega_{{\bf k+Q}}$ as we approach the ordering vector ${\bf Q}$. Similarly, the $|\langle 0 | S^z_{\bf k} | b \rangle |^2$ is also $\propto 1/\omega_{{\bf k}}$, however the divergency is now cut off by the anisotropy gap. In the absence of anisotropy, $S^{xx}({\bf k},\omega) = S^{zz}({\bf k},\omega)$, as expected. As for higher energy features,
the weights of the `c' and `d' modes in the spin response vanish with $\Lambda^2/J^2$ and $\Lambda^4/J^4$, and become negligible small for small values of $\Lambda$, as we already noticed in Fig.~\ref{fig:intensity}(a)-(c). 

It is also instructive to compare with the conventional spin-wave theory, where we keep the $b$ boson only (setting $\eta=1$ in Eq.~(\ref{eq:bboson}) is exactly the form of the spin operators using $b$ as the Holstein-Primakoff boson in the conventional case). The spin-wave Hamiltonian then reads 
\begin{eqnarray}
\mathcal{H}^{(2)} 
& = & 
S \left[
4 J - 2 \left(J - J_z\right)  \gamma_{\bf k} + \Lambda
\right]
\left(
  b_{\bf k}^{\dagger} b_{\bf k}^{\phantom{\dagger}} 
+ b_{\bf -k}^{\dagger} b_{\bf -k}^{\phantom{\dagger}} 
\right)
\nonumber\\&&
+ S \left[ 2 \left(J  + J_z\right) \gamma_{\bf k} - \Lambda \right] 
\left(
  b_{\bf k}^{\dagger} b_{\bf -k}^{\dagger} 
+ b_{\bf k}^{\phantom{\dagger}} b_{\bf -k}^{\phantom{\dagger}} 
 \right).
 \label{eq:Hb_1boson}
\end{eqnarray}
Replacing the $S=3/2$ value, the terms proportional to $J$ and $J_z$ are identical to the ones in Eq.~(\ref{eq:Hbd0}), while the coefficients of the terms involving $\Lambda$ differ.

\section{Modes for large single-ion anisotropy}
\label{sec:appendixB}

From Eq.~(\ref{eq:eta_sol}) we can obtain the $1/\Lambda$ expansion of $\eta$ in the
$\Lambda \gg J,J_z$ limit: 
\begin{equation}
 \eta = \frac{\Lambda }{3 J}-\frac{1}{3} + \frac{4J}{\Lambda} + O\left(\frac{J^2}{\Lambda^2}\right) \;.
\end{equation}
Diagonalizing the Hamiltonians (\ref{eq:Hbd}) and (\ref{eq:Hbd}) in this limit, we get the following excitation energies 
\begin{eqnarray}
 \omega^{b}_{\bf k} &=& 4 \sqrt{(1 - \gamma_{\bf k}) J \left(4 J + 
    J_z \gamma_{\bf k} +O(1/\Lambda) \right)} , \label{eq:wbL}\\
 \omega^{c}_{\bf k} &=& 2 \Lambda + 4 J - 3 J \gamma_{\bf k} + O(1/\Lambda), \\
 \omega^{d}_{\bf k} &=& 2 \Lambda + 4 J - 3 J \gamma_{\bf k} + O(1/\Lambda), 
\end{eqnarray}
and the matrix elements are  
\begin{eqnarray}
 |\langle 0 | S^x_{\bf k} | b \rangle |^2 &=& \frac{\sqrt{4 J - J_z \gamma_k}}{2\sqrt{J+J \gamma_k}} + O(1/\Lambda)\\
 |\langle 0 | S^z_{\bf k} | b \rangle |^2 &=& \frac{\sqrt{J-J \gamma_k}}{2\sqrt{4 J + J_z \gamma_k}} + O(1/\Lambda)\\
 |\langle 0 | S^y_{\bf k} | c \rangle |^2 &=& \frac{3}{4} - \frac{24+9 \gamma_k}{8}\frac{J}{\Lambda} + O(1/\Lambda^2)\\ 
 |\langle 0 | S^x_{\bf k} | d \rangle |^2 &=& \frac{3}{4} - \frac{24+9 \gamma_k}{8}\frac{J}{\Lambda} + O(1/\Lambda^2)\\
 |\langle 0 | S^z_{\bf k} | d \rangle |^2 &=& \frac{3J^2}{\Lambda^2}(1- \gamma_k)^2 + O(1/\Lambda^3) \;.
\end{eqnarray}
The weakly dispersing `c' and `d' modes describe excitations of the $|\pm 3/2\rangle$ $S_z$ states. We see that almost all the matrix elements are finite when the single-ion anisotropy is large. The smallnes of the $|\langle 0 | S^z_{\bf k} | d \rangle |^2$ can be understood as these are the fluctuations that need to overcome the large $\Lambda$ energy.
 
Here we can seek the correspondence to an effective model. When $\Lambda$ is large, the $S^z=\pm3/2$ states can be projected out from the Hamiltonian. The essential degrees of freedom are reduced to the $S^z=\pm1/2$ states of the $S=3/2$, and these two states and the interaction between them can be represented by the $\sigma^z=\pm 1/2$ states of an effective spin--1/2 model with the following XXZ Hamiltonian:
\begin{equation}
\mathcal{H}_{\text{eff}} = \sum_{<i,j>} \left( 4 J \left( \sigma_i^x \sigma_j^x + \sigma_i^y \sigma_j^y \right)+ J_z \sigma_i^z \sigma_j^z \right)
\end{equation}
 where the $\sigma^\alpha_i$ are spin-1/2 operators on site $i$ that act on the Hilbert space of the effective spins. So in the leading order we are left with the usual anisotropic antiferromagnetic spin-1/2 problem  that provides the same dispersion as $\omega_b$ in Eq.~(\ref{eq:wbL}).


\bibliographystyle{unsrt}
\bibliographystyle{apsrev4-1}
\bibliography{flavor_waves_v4_5}

\end{document}